\documentclass[fleqn,twoside]{article}
\usepackage{espcrc2}
\usepackage{amssymb}

\usepackage{graphicx}
\usepackage[figuresright]{rotating}

\newcommand{\be}{\begin{eqnarray}}
\newcommand{\ee}{\end{eqnarray}}

\newcommand{\bea}{\begin{eqnarray}}
\newcommand{\eea}{\end{eqnarray}}
\newcommand{\Xmax}{X_{\max}}

\newcommand{\AmS}{{\protect\the\textfont2
  A\kern-.1667em\lower.5ex\hbox{M}\kern-.125emS}}

\title{The QCD Black Disk Limit in Cosmic Ray Air Showers}

\author{H.~J.~Drescher\\
{\small\it Johann Wolfgang Goethe University, Postfach 11 19
  32, 60054 Frankfurt, Germany}\\
}

\begin{document}

\begin{abstract}
We discuss particle production in the high energy limit of QCD. Due to
a large gluon density, the interaction reaches the black disk limit
and the projectile is resolved into its partonic
structure at the saturation scale. This leads to suppression of
forward particle production and hereby to a faster absorption of
cosmic ray air showers. This property is most suitable for the
distinction of evolution scenarios for the saturation scale, e.g.
fixed and running coupling BFKL, the latter of which is favored by air
shower measurements. 
\end{abstract}

\maketitle

\section{Introduction}

The biggest uncertainty in air shower simulations is certainly the
hadronic interaction model, since QCD is poorly understood at
these high energies and accelerator data is not available. 

Higher twist corrections become increasingly important at high
energies. Attempts to account for this are  the
implementation of an  energy-dependent $p_t$ cutoff \cite{Sibyll} for
hard scattering or the resummation of enhanced pomeron diagrams in an
efficient manner \cite{ostap1}. This way one tries to extend the
applicability of hadronic interaction models up to GZK energies,
$E\approx 10^{11}$~GeV.

Our approach \cite{Drescher:2004sd} is to consider the black disk
limit (or black body limit - BBL) 
at high gluon densities within the Color Glass Condensate (CGC)
approach\cite{sat}, where the interaction probability is close to
unity. This leads to a suppression of forward particle scattering, the
most important phase space region for cosmic ray air showers. 

\section{Hadron nucleus scattering at very high energies}

\begin{figure}[tbh]
\includegraphics[width=\columnwidth]{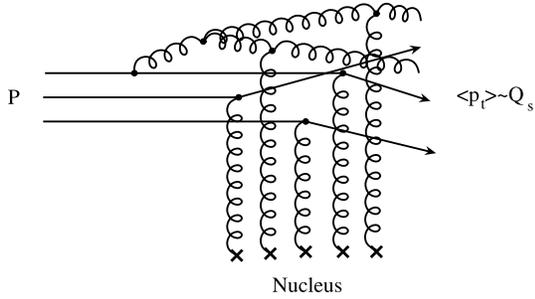}
\caption{Schematic view of the break-up of a hadron: the partonic
  structure is resolved at the saturation scale $Q_s$}
\label{fig:diag}
\end{figure}

The elastic and total scattering cross sections for quark-nucleus
scattering may be written as follows \cite{djm2}:
\be
\sigma^{\rm el} &=& \int d^2b \left[ 1-\exp(-Q_s^2/4\pi\Lambda^2)\right]^2\\
\sigma^{\rm tot} &=& 2\int d^2b \left[ 1-\exp(-Q_s^2/4\pi\Lambda^2)\right]~.
\ee
If $Q_s$ is large, the cross section approaches the geometrical limit.

Here, two different evolution scenarios are considered for the saturation
scale $Q_S$: fixed coupling
\be
Q_s^2(x,A)=Q_0^2(A)\left(\frac{x_0}{x}\right)^\lambda ~,
\ee
and running coupling \cite{IancuVenu}
\be
Q_s^2 &=& \Lambda^2 \exp(\log(Q_0^2/\Lambda^2)\sqrt{1+2c\alpha_s y}  ) \\
\alpha_s &=& \frac{b_0}{\log(Q^2/\Lambda^2)} ~,
\ee
with $y=\log(1/x)$. We assume $Q_0(A)$ to be proportional to the number of
participants $N_{\rm part}$. The constant $c$ assures a smooth
transition to the fixed coupling scenario at low $y$. At high energies
 $Q_s \approx 5$~GeV for running coupling and 20 GeV for
fixed coupling for a central p-N collision \cite{Drescher:2004sd}.

The diagrammatic structure of an BBL
event is schematically given in Fig. \ref{fig:diag}. The projectile interacts
with the target as a whole and loses its coherence\cite{DGS}. The partons are
resolved at a scale given by the saturation momentum. 
An important feature is that soft physics is mostly suppressed, the
typical transverse momentum being the resolution scale $Q_s$.

\section{Monte Carlo implementation}

\begin{figure}[t]
\includegraphics[width=\columnwidth]{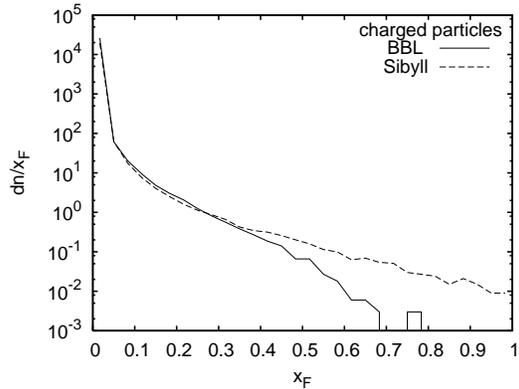}
\caption{ Suppression of forward scattering for central p-N collisions. }
\label{fig:xf0}
\end{figure}

\begin{figure}[t]
\includegraphics[width=\columnwidth]{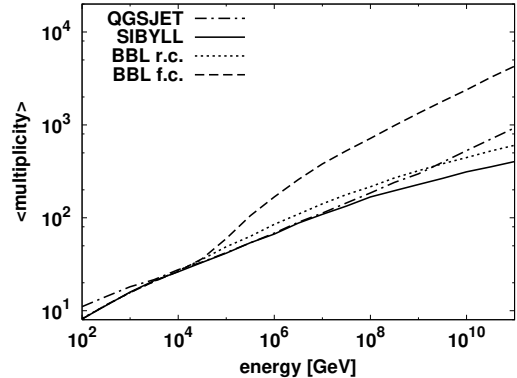}
\caption{ Mean multiplicity of central p-N events. }
\label{fig:mult}
\end{figure}

\begin{figure}[t]
\includegraphics[width=\columnwidth]{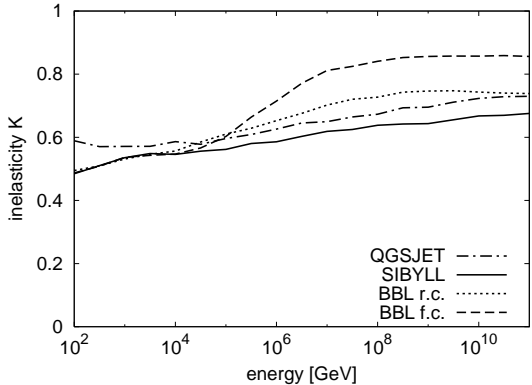}
\caption{Inelasticity as a function of energy.}
\label{fig:inel}
\end{figure}

\begin{figure}[t]
\includegraphics[width=\columnwidth]{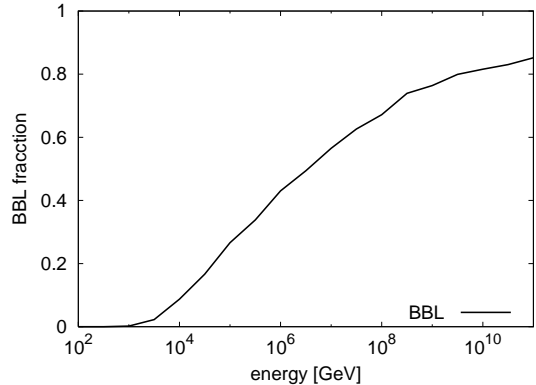}
\caption{ The fraction of BBL-events as a function of energy.}
\label{fig:bblfraction}
\end{figure}

Since the PDFs are resolved at the saturation scale $Q_s$, they
provide the probability distribution for a parton to appear in the
final state. The mean transverse momentum is approximately given by $Q_s$:
\be
P_i(x) &=& f_i\left( Q_s^2(x),x  \right) \label{for:P} \\
\left< p_t \right> &\approx& Q_s(x)
\ee
Note that  $Q_s$ is a function of $x$ and $b$. For valence quarks we use the
GRV94 parton distribution functions. For a baryon, first the three valence
quarks are generated, then the remaining energy is used for gluon
bremsstrahlung. Here we use the ansatz from Ref. \cite{KL}
\be
x g(x,q_t^2) \sim \frac{1}{\alpha_s} \min \left( q_t^2,Q_s^2(x)\right)(1-x)^2~,
\ee
which exhibits no divergence for low $q_t$. The produced gluons are
ordered in rapidity and placed on strings between the valence quarks
and the target nucleus, whose precise configuration is not important,
since we are interested in forward scattering. In principle, a
baryon-nucleus collision produces 3 strings, a meson-nucleus collision
2 strings. However, when the invariant mass between two of the three
quarks is small, one cannot  assume anymore that they fragment
independently. Therefore, we implement a cut-off in invariant mass,
\be
m_{\rm cut} = m_{\rho} = 0.77 {\rm~GeV}
\ee
below which  two leading quarks are allowed to form a diquark. This is an
important feature since it recovers the leading particle effect for
low $Q_s$. The hadronization of the strings is done within the LUND
model \cite{PY}. 

Of course, even at high energies, not all minimum bias events will be near
the BBL. Therefore, we couple our model to a standard pQCD leading twist
event generator, SIBYLL 2.1 \cite{Sibyll}. Which of the two models 
is to be applied for a given impact parameter, follows from the condition 
\be
Q^{\rm Nucleus}_s(x_F=0.001) \gtrsim 1 GeV  ~.
\ee
The saturation momentum of the target at a longitudinal momentum
fraction of $x=0.001$ for the projectile must be greater than a given
value. The typical resolution scale for the valence quarks is
correspondingly higher, since they have $x\approx 0.2$.

\section{Results of the BBL event generator}

Using the BBL as event generator, we can compare results in particle
production to the pQCD model SIBYLL. At the
same time, we compare the two different evolution scenarios, fixed and
running coupling.
Fig. \ref{fig:xf0} shows a Feynman-x distribution of charged particles
for central proton-nitrogen collisions. One notices the absence of particle
production in the very forward region, which is the typical
property of this approach.  This figure shows running coupling only;
fixed coupling would be even more extreme. 
The absence of forward scattering is compensated by a relatively
large multiplicity in the mid-rapidity region. In Fig. \ref{fig:mult},
we show the average multiplicities for fixed and running coupling BBL,
as well as SIBYLL and QGSJET01 \cite{qgsjet} results. Due to a missing
ad-hoc $p_t$ 
cutoff, QGSJET01 exhibits also a large multiplicity. 

An interesting observable for air showers is the inelasticity,
\be
K=1-\frac{E_{\max}}{E_0} ~,
\ee
where $E_{\max}$ is the energy of the most energetic particle.
A large value $K\approx 1$  means that most of the energy is used for
particle production. A small value $K\approx 0$ is significant for
elastic or diffractive events, where most of the energy remains in the
leading particle. For air showers, a high inelasticity means that the
shower are absorbed more rapidly in the atmosphere, giving rise to smaller
$\Xmax$ values. Fig. \ref{fig:inel} shows how the suppression of the
leading particle results in a higher $K$, especially for fixed
coupling.

Finally, we show the fraction of BBL-events as a function of primary
energy in Fig. \ref{fig:bblfraction}. 
It reaches 90\% at GZK energies. Note, that this does not mean
that 90\% of the cross section can be approximated as a black disk,
since this also depends on the $Q_s$ with which the event was
created. But it gives some idea about how this phenomenon becomes more
important at high energies. 

\section{Application of BBL to air showers}

\begin{figure}
\includegraphics[width=1.0\columnwidth]{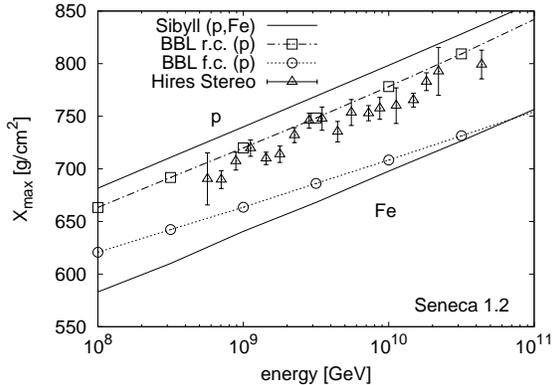}
\caption{\label{cap:Xmax} $X_{\max}$ as a function of primary energy.}
\end{figure}

We implement the BBL event generator into the Seneca air shower
code \cite{seneca} and compute the mean $\Xmax$ as a function of
energy for proton primaries. 

In Figure \ref{cap:Xmax} we compare the predictions of 
 SIBYLL~2.1 for proton and iron induced showers to the
saturation model (BBL, for proton primaries only) with running and
fixed coupling BFKL evolution of $Q_s$, respectively, and to
preliminary Hires stereo data~\cite{hires}. In the saturation limit,
showers do not penetrate as deeply into the atmosphere. This is due to
the ``break-up'' of the projectile's coherence~\cite{DGS} together with the
suppression of forward parton scattering (for central collisions).
The comparison to the data might suggest a preferably light
 composition, but the uncertainties at these energies are still
 considerable. 

\section{Conclusions}

We developed an hadronic interaction model considering the black disk
limit in high density QCD. The suppression of forward
scattering  results in  faster absorption of particles within
 air showers, which leads to a smaller $\Xmax$. 
This feature allows one to distinguish between possible
evolution scenarios for the saturation momentum. 

\section*{Acknowledgments}
This work was done in collaboration with A.~Dumitru and M.~Strikman.

H.-J.D.~acknowledges support by the German Minister for
Education and Research (BMBF) under project DESY 05CT2RFA/7.
The computations were performed at the
 Frankfurt Center for Scientific Computing (CSC).

\end{document}